\documentclass[%
 reprint,
 amsmath,amssymb,
 aps,
]{revtex4-2}

\usepackage{enumerate}
\usepackage{amsmath}
\usepackage{graphicx}
\usepackage{dcolumn}
\usepackage{bm}

\begin{document}

\preprint{APS/123-QED}

\title{How the zebra got its stripes:\\Curvature-dependent diffusion orients Turing patterns on 3D surfaces}

\author{Michael F. Staddon}
\affiliation{Center for Systems Biology Dresden, Dresden, Germany}
\affiliation{Max Planck Institute for the Physics of Complex Systems, Dresden, Germany}
\affiliation{Max Planck Institute of Molecular Cell Biology and Genetics, Dresden, Germany}

\date{\today}

\begin{abstract}
Many animals have patterned fur, feathers, or scales, such as the stripes of a zebra.
Turing models, or reaction-diffusion systems, are a class of mathematical models of interacting species that have been successfully used to generate animal-like patterns for many species. When diffusion of the inhibitor is high enough relative to the activator, a diffusion-driven instability can spontaneously form patterns.
However, it is not just the type of pattern but also the orientation that matters, and it remains unclear how this is done in practice.
Here, we propose a mechanism by which the curvature of the surface influence the rate of diffusion, and can recapture the correct orientation of stripes on models of a zebra and of a cat in numerical simulations.
Previous work has shown how anisotropic diffusion can give stripe forming reaction-diffusion systems a bias in orientation. From the observation that zebra stripes run around the direction of highest curvature, that is around the torso and legs, we apply this result by modifying the diffusion rate in a direction based on the local curvature.
These results show how local geometry can influence the reaction dynamics to give robust, global-scale patterns.
Overall, this model proposes a coupling between the system geometry and reaction-diffusion dynamics that can give global control over the patterning by using only local curvature information. Such a model can give shape and positioning information in animal development without the need for spatially dependent morphogen gradients.

\end{abstract}

\maketitle


\section{Introduction}


Many animals have patterned fur, feathers, or scales that can serve many different purposes such as the camouflaging stripes of a tiger or zebra, sexual selection as in the peacocks feathers, or as a warning signal in striped venomous snakes or caterpillars. For striped patterns, it can be important that the stripes are oriented in a specific manner for effective function. A tiger with horizontal stripes might stick out from the vertical blades of grass and fail to sneak up on its prey. In fact, the orientation of the stripes depends on the location on the body; for a standing tiger or zebra, stripes are aligned vertically around the torso and horizontally around the legs, and it remains unclear how these patterns are arranged during development.

Alan Turing famously proposed a model for pattern formation~\cite{turing1990chemical}, now known as Turing models or reaction-diffusion models. Typically there are two interacting species, one known as the activator, which increases production of the two species, and the other the inhibitor which decreases production. When the activator is strong enough, and the inhibitor diffuses fast enough, then a diffusion-driven instability can produce periodic patterns from an initially uniform state, known as Turing patterns~\cite{turing1990chemical, murray1973mathematical, kondo2010reaction, marcon2012turing}. Depending on the interactions one can obtain either spots or stripes, and there has been theoretical success in reproducing a wide range of animal pigmentation patterns~\cite{kondo1995reaction, barrio1999two, nakamasu2009interactions, fofonjka2021reaction, yang2023computer}, digit formation in hands and feet~\cite{sheth2012hox, maini1992pattern}, and hair and feather positioning~\cite{jung1998local, sick2006wnt}. More recently there has also been experimental evidence of biological circuits that could act as Turing models~\cite{kondo1995reaction, nakamasu2009interactions, muller2012differential, sick2006wnt, jung1998local}.

Standard Turing models generate patterns with no bias in orientation, but in nature it is often important that patterns are aligned in the correct direction. One argument is that the geometry of the animal may be enough to limit waves in certain directions, for example, the wave length of stripes on a tigers tail may be larger than the tails circumference, and so patterns may only emerge with stripes going around the tail~\cite{murray1973mathematical, lacalli1988theoretical}. However, this argument clearly breaks down when looking at the animals torso which often have vertical stripes. Instead, there are several additional mechanisms that can orient Turing-like patterns~\cite{hiscock2015orientation}. Including a spatial gradient of chemical source for the activator can give stripes aligned perpendicular to the gradient~\cite{barrio1999two, glimm2012reaction}. Similarly, a spatial gradient in the interaction parameters can give stripes aligned with the gradient~\cite{sheth2012hox, yang2023computer}. Alternatively, anisotropy in the relative diffusion between activator or inhibitor~\cite{shoji2002directionality, busiello2015pattern} or anisotropic growth of the domain~\cite{krause2019influence} can also lead to stripe orientation.

From the observation that stripes of cats and zebras typically follow the direction of curvature, vertically around the torso and horizontally around the legs when standing, we propose a model for pattern alignment in which the diffusion is coupled to the curvature of the surface. First, we rederive the conditions for pattern formation in reaction-diffusion models and show that stripes grow fastest in the direction of highest relative diffusion between inhibitor and activator. Next we perform numerical simulations on periodic 2D domains with anisotropic diffusion and show that even for small differences in diffusion we obtain robust alignment of stripes. In the final section we perform numerical simulations on 3D models of cats and zebras and show that with curvature coupling we can obtain patterns qualitatively similar to those in nature. Finally, we discuss potential biological mechanisms for curvature-coupled diffusion. Overall, this proposed coupling gives a simple mechanism for pattern alignment as seen in nature using only local properties to give robust global pattern formation.

\section{Instability and wavelengths increase with inhibitor diffusion}

In this first section, we consider a generic 2-species reaction-diffusion system~\cite{turing1990chemical}. Starting from a uniform steady state with a small perturbation, patterns spontaneously form as these perturbations grow when the Turing conditions hold. We show that increasing the diffusion coefficient of the inhibitor always increases the growth rate and wavelength of these instabilities.

Consider a general reaction diffusion system with isotropic diffusion
\begin{equation}
    \dot{u} = D_u \nabla^2 u + f(u, v)
\end{equation}
and
\begin{equation}
    \dot{v} = D_v \nabla^2 v + g(u, v).
\end{equation}
where $f$ and $g$ are the chemical reaction rates, and $D_u$ and $D_v$ are diffusion rates of $u$ and $v$. Without loss of generality, we may simplify our equations to
\begin{equation}
    \dot{u} = \nabla^2 u + f(u, v)
\end{equation}
and
\begin{equation}
    \dot{v} = d \nabla^2 v + g(u, v)
\end{equation}
which can be achieved by rescaling the $x$ and $y$ coordinates.

The system has steady states $u = u^*$ and $v = v^*$ such that $f(u^*, v^*) = 0$ and $g(u^*, v^*) = 0$. By considering small perturbations to this steady state of the form $u = u_* + \tilde{u}e^{i(k_x x + k_y y)}$ and $v = v_* + \tilde{v}e^{i(k_x x + k_y y)}$ and using linear stability analysis, one can show that the conditions for Turing instabilities are:
\begin{enumerate}[(i)]
    \item $f_u + g_v < 0$
    \item $f_u g_v - f_v g_u > 0$
    \item $d f_u + g_v > 0$
    \item $(d f_u + g_v)^2 - 4d(f_u g_v - f_v g_u) > 0$
\end{enumerate}
where $f_u = \partial_u f$ and so on (see the supplementary materials for a full derivation). The first two conditions are for a steady uniform state and the last two conditions are for a diffusion-drive instability. A corollary of conditions (i) and (iii) is that $d \neq 1$, and further, if $f_u > 0 \implies d > 1$ and $f_u < 0 \implies d < 1$, which means we must have a relatively slow diffusing activator and fast moving inhibitor for patterns to form.

When these conditions hold, in the case where $u$ is the activator with $f_u > 0$, we find that the maximum growth rate across all frequencies $k = \sqrt{k_x^2 + k_y^2}$ is given by
\begin{equation}
    \lambda(k_*) = \frac{1}{2} \left( (f_u + g_v) + \frac{d+1}{d-1} (f_u - g_v) - \frac{4d}{d-1} \sqrt{-\frac{f_v g_u}{d} } \right)
\end{equation}
with the corresponding frequency
\begin{equation}
    k_* = \sqrt{-\frac{f_u - g_v}{d-1} + \frac{d+1}{d-1} \sqrt{-\frac{f_v g_u}{d}}}.
\end{equation}

Next, we show that increasing diffusion rate of the inhibitor always increases both the wavelength and growth rate of the instability. The derivative of growth rate with respect to diffusion coefficient $d$ is given by
\begin{equation}
\partial_d \lambda(k_*) = \frac{1}{2(d-1)^2} \left(-(f_u - g_v) + (d+1) \sqrt{-\frac{f_v g_u}{d} } \right),
\end{equation}
which is always positive since by it is proportional to $k_*^2 > 0$, in the case that $f_u > 0$. Similarly, when $f_u < 0$ the derivative is negative, meaning that increasing the diffusion of the inhibitor against the activator always increases the instability. Further, for $d$ large enough, the frequency $k_*$ always decreases, meaning that further increases to the diffusion rate also increase the pattern wavelength.

\begin{figure}[h]
\includegraphics[width=0.5\textwidth]{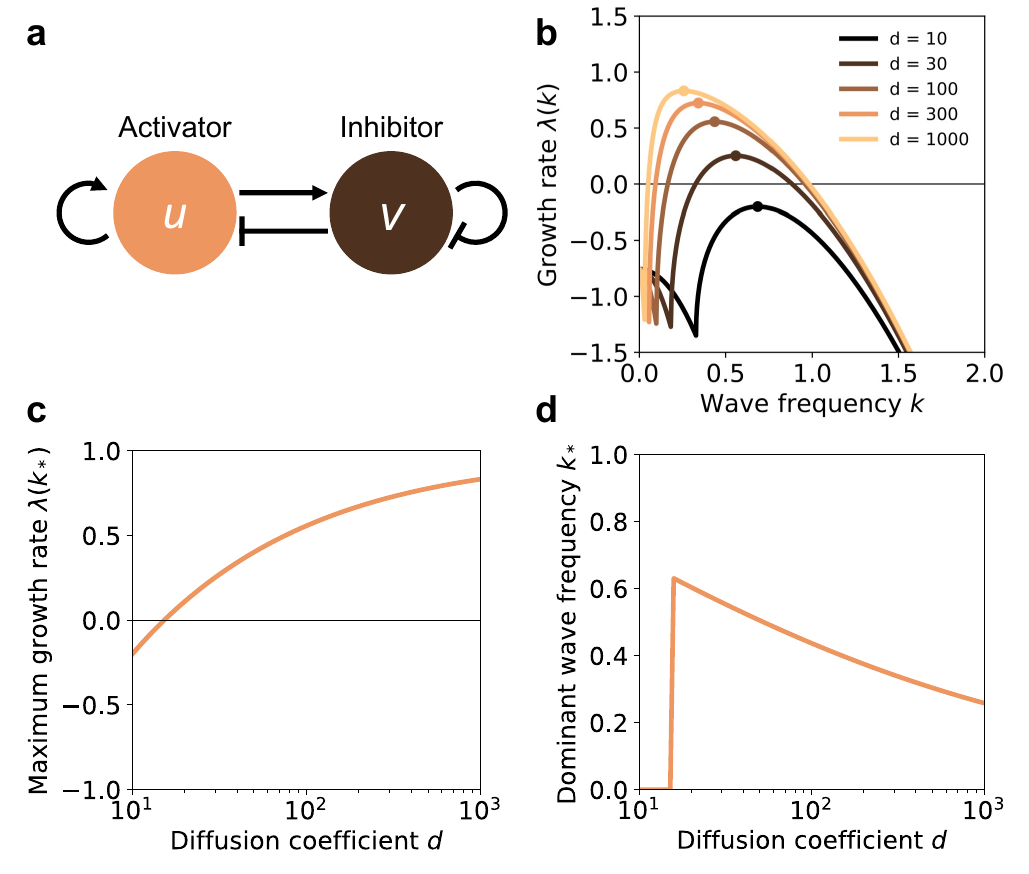}
\caption{Inhibitor diffusion increases instabilities. \textbf{a}, Example 2-species reaction diffusion system. The activator $u$ increases production of both $u$ and $v$ while the inhibitor $v$ decreases production. \textbf{b} Instability growth rate $\lambda$ against wave number $k$ for different diffusion coefficients $d$. \textbf{c}, Maximum growth rate across all wave numbers against diffusion coefficient $d$. $\textbf{d}$, The dominant wave number $k_*$, which maximises the growth rate, against diffusion coefficient $d$.}
\centering
\label{fig:1}
\end{figure}

To demonstrate these results with an example, we use the Schnakenberg model~\cite{schnakenberg1979simple}, originally describing the kinetics of glycolosis, which can produce striped or spotted patterns depending on the reaction parameters. The rate equations are given by
\begin{equation}
    \dot{u} = \nabla^2 u + a +  u^2 v - u
\end{equation}
\begin{equation}
    \dot{v} = d \nabla^2 v + b - u^2 v
\end{equation}
such that $f(u, v) = a +  u^2 v - u$ and $g(u, v) = b - u^2 v$, and has steady state given by $u_* = a + b$ and $v_* = \frac{b}{(a + b)^2}$. In this instance, $u$ is the activator and $v$ the inhibitor (Fig.~\ref{fig:1}a). We calculate the growth rate, using $a = 0.025$ and $b = 1.55$, and find that diffusion coefficient for the inhibitor must be higher than that of the activator, as expected by condition (iii), with the growth rate of the instabilities increasing with $d$ (Fig.~\ref{fig:1}b, c). At the same time, the most unstable wave frequency decreases, meaning the pattern wavelength increases as diffusion increases (Fig.~\ref{fig:1}d).

\section{Anistropic diffusion aligns Turing patterns}

Using the result that increasing diffusion of the inhibitor increases the instabilities growth rate, it follows that for anistropic diffusion, the instability will grow fastest in the direction of highest inhibitor diffusion. Consider a reaction diffusion system with anisotropic diffusion
\begin{equation}
    \dot{u} = \partial_x^2 u + \partial_y^2 u + f(u, v)
\end{equation}
and
\begin{equation}
    \dot{v} = d_x \partial_x^2 v + d_y \partial_y^2 v + g(u, v).
\end{equation} 
To find the most unstable wavelength, we write our wave vector in polar coordinates as $(k, l) = (r \cos \theta, r \sin \theta)$. This converts our equations for this perturbation into the form discussed in the section above, with our effective diffusion constant $d(\theta) = d_x \cos^2 \theta + d_y \sin^2 \theta$. Following from the final result, when the system is unstable, the growth rate will be highest in the direction of greatest diffusion if $u$ is an activator, or in the direction of lowest diffusion if $u$ is an inhibitor. In addition, it is possible for the perturbation to be stable in one direction and unstable perpendicular to it~\cite{busiello2015pattern}. This means that in cases where we have anisotropic diffusion, we expect stripes to align either parallel or perpendicular to the direction of highest relative diffusion, that is, the highest ratio between the diffusion coefficients in the $x$ direction or $y$ directions.

To test these predictions, we simulate the Schnakenberg equations in a periodic 2-dimensional box with anisotropic diffusion in the inhibitor:
\begin{equation}
    \dot{u} = \partial_x^2 u + \partial_y^2 u + f(u, v)
\end{equation}
\begin{equation}
    \dot{v} = d_x \partial_x^2 v + d_y \partial_y^2 v + g(u, v).
\end{equation}
We set $d_y = 20$ and vary diffusion in the $x$ direction from $d_x = 0.5 d_y$ to $d_x = 2 d_y$. Starting from the uniform steady state with $a = 0.025$ and $b = 1.55$, parameters which would give stripes when $d_x = d_y = 20$, we add a small normally distributed perturbation and evolve the system numerical until a steady state is reached. Numerical details are contained in the supplement.

\begin{figure}[h]
\includegraphics[width=0.5\textwidth]{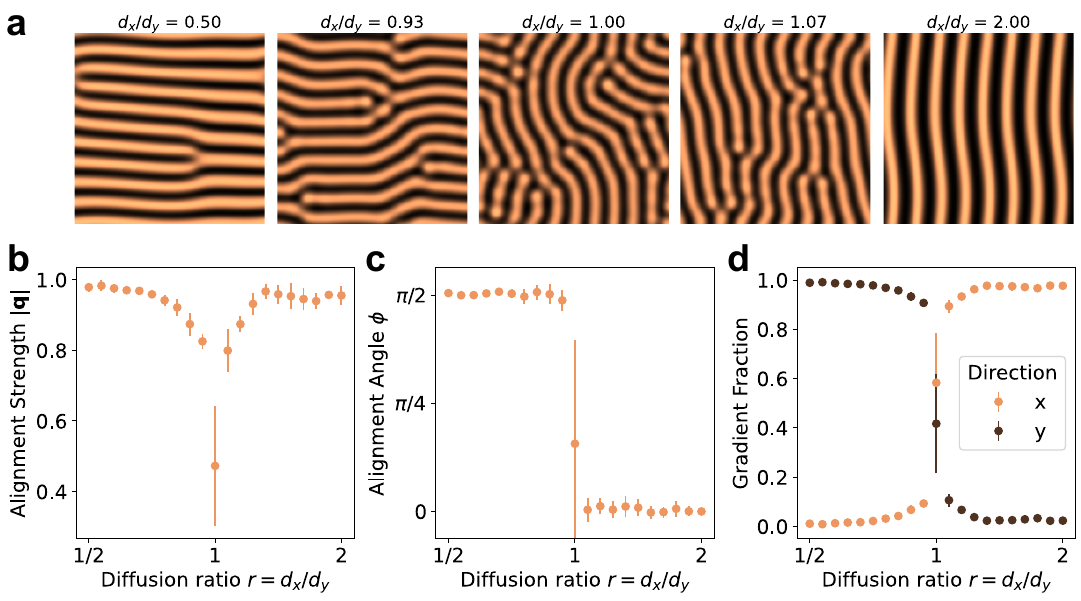}
\caption{Anisotropic diffusion gives robust pattern alignment. \textbf{a}, Simulation results of the Schnkackenberg equations for varying ratios of $d_x / d_y$, where $d_y = 20$. \textbf{b}, magnitude $|\mathbf{\bar{q}}|$ and $\textbf{c}$ orientation angle $\phi$, of the average gradient nematic tensor $\mathbf{\bar{q}}$ against the diffusion ratio $r = d_x / d_y$. \textbf{d}, Gradient fraction in the $x$ and $y$ directions against the diffusion ratio. Dots show the mean and standard deviation (n = 10).}
\centering
\label{fig:2}
\end{figure}

When $d_x = d_y$ we obtain striped patterns with no bias in orientation (Fig.~\ref{fig:2}a, middle panel). When $d_x < d_y$, we find that stripes align with the $x$ direction, even for differences in diffusion as small as $d_x = 0.93 d_y$ (Fig.~\ref{fig:2}a, left panels). Similarly, when $d_x > d_y$ we see stripes aligning with the $y$ direction instead (Fig.~\ref{fig:2}a, right panel).

We quantify pattern alignment by analysing the nematic tensor formed by the gradient of $u$:
\begin{equation}
    \mathbf{q} = 2 \nabla u \otimes \nabla u / |\nabla u|^2 - \mathbb{I}.
\end{equation}
For each simulation we average the nematic tensor over space, weighted by the square of the gradient
\begin{equation}
    \mathbf{\bar{q}} = \frac{\int \mathbf{q} |\nabla u|^2 dA}{\int |\nabla u|^2 dA}.
\end{equation}
The averaged tensor for each simulation is of the form
\begin{equation}
    \mathbf{q} = |\mathbf{q}| \begin{pmatrix}
        \cos 2\phi & \sin 2\phi \\
        \sin 2 \phi & -\cos 2\phi
    \end{pmatrix}
\end{equation}
where $|\mathbf{q}|$ is the alignment strength and $\phi$ is the alignment angle, which shows the preferred direction of the gradient and is perpendicular to the direction of the stripes. When $d_x = d_y$ there is low alignment strength $|\mathbf{q}|$ (Fig.~\ref{fig:2}b) with no preferred orientation angle $\phi$, as shown by the large variance in values across simulations (Fig.~\ref{fig:2}c). However, for $d_x \neq d_y$ we find the alignment strength quickly increases towards 1 as the difference between $q_x$ and $q_y$ increases (Fig.~\ref{fig:2}d), with a strong preference for an alignment angle equal to $\pi / 2$ for $d_x < d_y$ with little variance across simulations, giving horizontal stripes, and for and alignment angle around 0 for $d_x > d_y$, giving vertical stripes (Fig.~\ref{fig:2}c). These results indicate that even small changes in diffusion in given directions can give a strong preference for pattern orientation.

As an alternative metric for alignment, we may define the gradient-fraction in the $x$ direction as
\begin{equation}
    G_x = \frac{\int \partial_x u^2 dA}{\int (\partial_x u ^ 2 + \partial_y u ^2) dA}
\end{equation}
and the $y$ direction as
\begin{equation}
    G_y = \frac{\int \partial_y u^2 dA}{\int (\partial_x u ^ 2 + \partial_y u ^2) dA}
\end{equation}
which effectively measures what fraction of the gradient lies in a given direction. While not as informative as the nematic tensor analysis, it gives an intuitive understanding of the pattern and a similar method will be used in the last section of this paper on 3D models. Similar to the previous analysis, for $d_x = d_y$ there is no clear preference in pattern direction. However, for $d_x < d_y$ most of the gradient is in the $y$ direction, meaning horizontal stripes, and for $d_x > d_y$ most of the gradient is in the $x$ direction (Fig.~\ref{fig:2}d).

\section{Curvature-diffusion coupling reproduces animal patterns}

In this section, we introduce the idea of curvature-diffusion coupling and show how it can predict both pattern alignment and wavelengths. We have seen in the previous section how small changes in diffusion with direction can align stripes. Here, we allow diffusion to be modified by the curvature of a 3D surface and find that we can qualitatively reproduce patterns observed striped in animals such as cats and zebras, where stripes run around the torso and legs, which are locally in the direction of highest curvature.

On a 3D surface, we have the surface normal $\mathbf{n}$, and two directions of principal directions $\mathbf{t}_1$ and $\mathbf{t}_2$, with curvatures $\kappa_1$ and $\kappa_2$ respectively. We modify the diffusion of the inhibitor $v$ in to be anisotropic and curvature dependent, giving equations
\begin{equation}
    \dot{u} = \nabla^2 u + f(u, v)
\end{equation}
\begin{equation}
    \dot{v} = \nabla \cdot \left[ (d(k_1) \mathbf{t}_1 \otimes \mathbf{t}_1 + d(k_2) \mathbf{t}_2 \otimes \mathbf{t}_2) \cdot \nabla v \right] + g(u, v)
\end{equation}
where $d(\kappa)$ is a curvature dependent diffusion coefficient. This diffusion term means that when the gradient of $v$ is only in the $\mathbf{t}_i$ direction then it diffuses with rate $d(\kappa_i)$.

We choose a simple monotonic diffusion coefficient in the form of a logistic function
\begin{equation}
    d(\kappa) = d_0 \left(\frac{1}{2} + \frac{1}{1 + e^{-\kappa h_d}}\right)
\end{equation}
where $d_0$ is the diffusion coefficient at zero curvature, and $h_d$ is the curvature-diffusion coupling strength. When $h_d$ is positive, then increasing curvature increases the diffusion coefficient and we should expect stripes to align perpendicular to the direction of curvature. In contrast, when $h_d$ is negative then diffusion decreases with curvature, and so stripes should align parallel to the direction of curvature. One consequence of this coupling is that the stripe width will also depend on curvature, since increasing diffusion increases the pattern wave length (Fig.~\ref{fig:1}d). For negative coupling we should expect smaller stripes around highly curved regions such as the legs and tail when compared to less curved regions such as the torso.

We simulate curvature-coupled Schnakenberg equations on a mesh of a cat and on a mesh of a horse. The surface curvature is calculated using the pymeshlab package in python, and simulations are numerically integrated using the fipy package, with details given in the supplement. Starting from a uniform steady state plus some noise, we evolve the equations until a steady state pattern is reached.

\begin{figure}[h]
\includegraphics[width=0.5\textwidth]{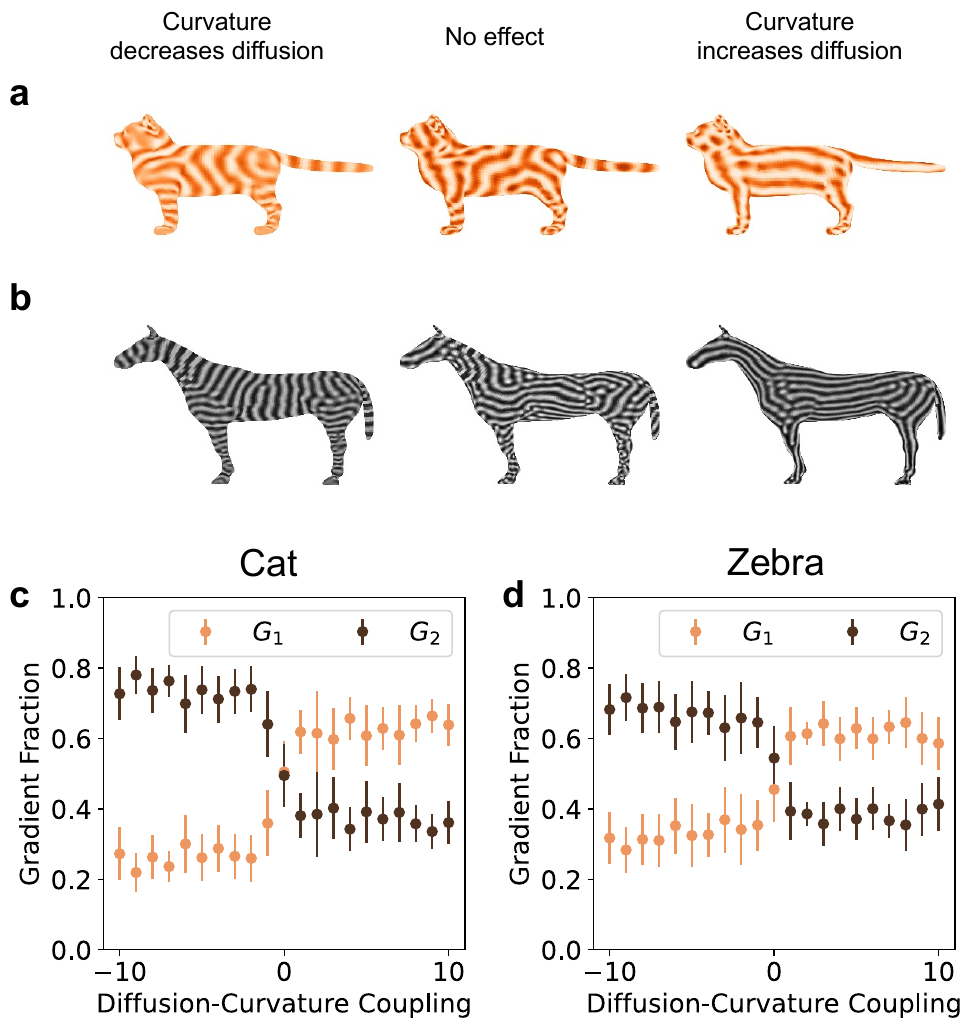}
\caption{Diffusion coupled to curvature gives robust alignment on curved surfaces. \textbf{a - b}, Simulation results of the Schnakenberg equations on 3D models of \textbf{a}, a cat and \textbf{b}, a zebra for different curvature-diffusion coupling strengths. \textbf{c - d}, Gradient fraction in directions of principle curvature, $\mathbf{t}_1$ and $\mathbf{t}_2$ for $\mathbf{t}_1$, for \textbf{c} cats and \textbf{d} zebras. Dots show the mean and standard deviation over 10 simulations.}
\centering
\label{fig:3}
\end{figure}

For no curvature coupling, $h_d = 0$, we observe stripe formation with no clear preferred orientation (Fig.~\ref{fig:3}a-b). In particular, along the torso and the legs we see no clear orientation. On the horses tail, the stripes align with the curvature, likely due to the radius of curvature being smaller than the typical wave size, and so it is unable to form a gradient in that direction.

For negative curvature coupling, $h_d < 0$, such that the diffusion coefficient decreases with curvature we find patterns similar to those observed in tigers and zebras (Fig.~\ref{fig:3}a-b). Along the neck and torso of the animals the stripes are vertical, since they align with the curvature. Along the the legs the stripes are instead horizontal and have a smaller wavelength.

For positive curvature coupling, $h_d > 0$, we see patterns aligning perpendicular to the direction of highest curvature, resulting in a pattern with horizontal stripes along its torso, more like a zebrafish than a zebra (Fig.~\ref{fig:3}a-b). Additionally, despite the small circumference of the tail, a single horizontal stripe is able to form, running the length of the tail.

To quantify the effect of curvature coupling, we perform a similar analysis to the 2D case, in which we define the gradient fraction in the direction of principal curvatures, $\mathbf{t}_i$, by
\begin{equation}
    G_i = \frac{\int (\mathbf{t}_i \cdot \nabla u)^2 dS}{\int (\nabla u)^2 dS}.
\end{equation}
For both the cat and zebra model, we find that for negative coupling, $h_d < 0$ the gradient mostly align perpendicular to the direction of highest curvature, $\mathbf{t}_1$, giving stripes aligned with curvature (Fig.~\ref{fig:3}c-d). As $h_d$ is increased the coupling remains relatively strong until close to zero coupling, at which point the orientation appears to take no bias. For positive coupling, we instead see most of the gradient in the $\mathbf{t}_1$ direction, giving stripes perpendicular to the direction of highest curvature.

\section{Discussion}

In this paper, we have shown how reaction-diffusion systems that form stripes can be oriented by including anisotropic diffusion (Fig.~\ref{fig:1}). When the inhibitor diffuses faster in one direction, relative to the activator, then stripes will be formed perpendicular to that direction, with alignment occurring even for small differences in diffusion (Fig.~\ref{fig:2}). Finally, by including diffusion which is influenced by surface curvature, we qualitatively capture the striped patterns observed in many animals (Fig.~\ref{fig:3}). When curvature decreases diffusion, stripes align around the torso and legs, as observed in zebras and cats. Moreover, a consequence of the decreased diffusion around the legs, relative to the torso, is the reduction of stripe width which is also seen in these animals. In contrast, when curvature increases diffusion, stripes run along the torso and down the legs.

This model proposes a mechanism for global pattern organisation using only knowledge of the local curvature. This stands in contrast with work by Yang and Kim~\cite{yang2023computer}, which generates accurate patterns of zebra stripes but uses spatially varying patterns to produce these. The patterns generated here are however not perfect. The heads of many striped animals often include many twists and turns of the patterns, while in this work we get concentric rings of stripes around the head for the zebra~\ref{fig:3}. However, the head also includes many topological features not included in the simple 3D meshes used here. The eyes, ears, and mouth all act as holes on the skin surface, which could add boundary conditions to the pattern and prevent diffusion across them. A more accurate mesh may also included finer details of curvature on the head which would further influence the results.

While the coupling between diffusion and curvature is purely theoretical, there are several biological candidates that could enable a similar coupling. A simple idea would be that the activator and inhibitor diffuse on slightly offset surfaces, for example on the apical and basal sides of a cell, or through a bilayer of cells. When the top surface curves, the bottom surface must change its length by a different amount over the same angle, and thus would speed or slow diffusion relative to the other. Alternatively cells may sense and respond to the curvature, modifying the reaction-diffusion parameters in response~\cite{callens2020substrate, luciano2021cell, tang2022collective}. In the case where the activator and inhibitor are different cell types, then their crawl speeds may be curvature dependent which would affect the effective diffusion rate~\cite{song2015sinusoidal, pieuchot2018curvotaxis}. Finally, cells may physically be stretched in certain directions by the curvature of the surface and have anisotropic shapes, for example in the many developing birds~\cite{curantz2022cell}. If diffusion of the morphogen is limited by cell boundaries, then having less cells per unit length, due to strain induced by the curvature, would give an increase the effective diffusion rate. In contrast, if diffusion is limited by the cytoplasm then straining cells has no effect on the effective diffusion rate.

Overall, this work proposes a new coupling between reaction-diffusion models and the system geometry, and applies this to striped pattern formation and orientation. An interesting follow up question would be to study the case where the curvature is influenced by the reaction-diffusion system, for example by activator induced growth. In this case, a feedback loop would exist between the reaction-diffusion system and the system geometry that may result in complex behaviour from a simple ruleset.

\begin{acknowledgments}
I thank my late cat Snoop whose stripes inspired this work. I thank Pierre Haas and Carl D. Modes for proof reading the manuscript and giving valuable feedback.

\end{acknowledgments}

\nocite{*}

\bibliography{refs}

\end{document}


\preprint{APS/123-QED}

\title{Supplementary Material\\How the tiger got its stripes:\\Curvature-diffusion coupling orients Turing patterns on 3D surfaces}

\author{Michael F. Staddon}
\affiliation{Center for Systems Biology Dresden, Dresden, Germany}
\affiliation{Max Planck Institute for the Physics of Complex Systems, Dresden, Germany}
\affiliation{Max Planck Institute of Molecular Cell Biology and Genetics, Dresden, Germany}

\date{\today}

\maketitle

\section{Derivation of Turing Conditions}
The system has steady states $u = u^*$ and $v = v^*$ such that $f(u^*, v^*) = 0$ and $g(u^*, v^*) = 0$. A small perturbation to a uniform steady state of the form $u = u_* + \tilde{u}e^{i(k_x x + k_y y)}$ and $v = v_* + \tilde{v}e^{i(k_x x + l_y y)}$ evolves as
\begin{equation}
    \begin{pmatrix}
    \partial_t \tilde{u} \\
    \partial_t \tilde{v}
    \end{pmatrix}
    = 
    \begin{pmatrix}
    -k^2 + f_u & f_v \\
    g_u                & - d k^2 + g_v
    \end{pmatrix}
    \cdot
    \begin{pmatrix}
        \tilde{u} \\
        \tilde{v}
    \end{pmatrix} 
\end{equation}
where
\begin{equation}
    J = \begin{pmatrix}
    -k^2 + f_u & f_v \\
    g_u                & - d k^2 + g_v
    \end{pmatrix}
\end{equation}
is the Jacobian, and $k = \sqrt{k_x^2 + k_y^2}$ is the wave frequency.

This system of equations has eigenvalues
\begin{equation}
    \lambda_{\pm} = \frac{1}{2}\left(  T \pm \sqrt{T^2 - 4 D} \right)
\end{equation}
where
\begin{equation}
    T(r) = (f_u + g_v) - (1 + d) k^2
\end{equation}
is the trace of the Jacobian matrix $J$ and
\begin{equation}
    D(r) = (k^2 - f_u)(d k^2 - g_v) - f_v g_u
\end{equation}
is the determinant. If one of the eigenvalues is positive then the perturbation grows over time and patterns spontaneously form.

For $u_*$ and $v_*$ to be a uniform steady state we must have
\begin{equation}
    T(0) < 0 \implies f_u + g_v < 0
\end{equation}
and
\begin{equation}
    D(0) > 0 \implies f_u g_v - f_v g_u > 0
\end{equation}
so that any spatially uniform perturbations have a negative growth rate. For a non-uniform state with wave number $k$ to be unstable, we require a positive $\lambda$. Since from above $T < 0$ for $k = 0$ and it is a decreasing function of $k$, we require that $D < 0$ for some $k$. Since $D$ is quadratic in $k^2$ we require that the second term is negative, meaning
\begin{equation}
    d f_u + g_v > 0
\end{equation}
otherwise all terms would be positive.

Finally, we can find the minimum of $D$ which must be below zero: 
\begin{equation}
    \underset{k}{min} D = \frac{1}{4d} (4d(f_u g_v - f_v g_u) - (d f_u + g_v)^2)
\end{equation}
and so we have the condition
\begin{equation}
    (d f_u + g_v)^2 - 4d(f_u g_v - f_v g_u) > 0.   
\end{equation}

These four conditions are the conditions for Turing instabilities~\cite{turing1990chemical, busiello2015pattern}:
\begin{enumerate}[(i)]
    \item $f_u + g_v < 0$
    \item $f_u g_v - f_v g_u > 0$
    \item $d f_u + g_v > 0$
    \item $(d f_u + g_v)^2 - 4d(f_u g_v - f_v g_u) > 0$
\end{enumerate}
The first are the conditions for a steady uniform state and the last two are the conditions for diffusion-drive instability. A corollary of conditions (i) and (iii) is that $d \neq 1$, and further, if $f_u > 0 \implies d > 1$ and $f_u < 0 \implies d < 1$, which means we must have a relatively slow diffusing activator and fast moving inhibitor for patterns to form.

Next, we determine the maximum growth rate across all wave frequencies $k$. The largest eigenvalue for a given $k$ is
\begin{equation}
\begin{split}
     \lambda = &\frac{1}{2}\left((f_u + g_v) - (d + 1)k^2 \right) + \\
        &\frac{1}{2}\sqrt{((d - 1) k^2 + (f_u - g_v))^2 + 4 f_v g_u}
\end{split}
\end{equation}
with derivative
\begin{equation}
    \partial_k \lambda = -(d+1)k + \frac{(d-1)((d-1)k^2 + (f_u - g_v))k}{\sqrt{((d - 1) k^2 + (f_u - g_v))^2 + 4 f_v g_u}}.
\end{equation}
This has fixed point at
\begin{equation}
    k =  \sqrt{- \frac{f_u - g_v}{d-1} \pm \frac{d+1}{d-1} \sqrt{-\frac{f_v g_u}{d}}}.
\end{equation}
Since we require $k$ to be real, for $f_u > 0$ we have
\begin{equation}
    k_{+} = \sqrt{-\frac{f_u - g_v}{d-1} + \frac{d+1}{d-1} \sqrt{-\frac{f_v g_u}{d}}}
\end{equation}
and for $f_u < 0$ we have
\begin{equation}
    k_{-} = \sqrt{-\frac{f_u - g_v}{d-1} - \frac{d+1}{d-1} \sqrt{-\frac{f_v g_u}{d}}}.
\end{equation}
When $f_u < 0$ then we require that $d < 1$ and $k_-$ gives the highest growth rate, and when $f_u > 0$ we require $d > 1$ and $k_+$ gives the highest growth rate. In both cases we require the wave number $k$ to be real and have a fifth condition:
\begin{enumerate}[(i)]
    \setcounter{enumi}{4}
    \item $(f_u - g_v)^2 + \frac{(d+1)^2}{d} f_v g_u < 0$.
\end{enumerate}
The maximum eigenvalue is given by
\begin{equation}
    \lambda(k_\pm) = \frac{1}{2} \left( (f_u + g_v) + \frac{d+1}{d-1} (f_u - g_v) \mp \frac{4d}{d-1} \sqrt{-\frac{f_v g_u}{d} } \right)
\end{equation}
with the choice of $k_{\pm}$ determined by the signs of $f_u$ and $g_v$.

\section{2D Simulations}

All numerical simulations were performed using the fipy python package~\cite{Guyer_Wheeler_Warren_2009}, a finite volume PDE solver. We simulate the equations in a 2D periodic box with side lengths 100, with a 100 grid points in each direction. Starting from a uniform steady state, we apply a normally distributed perturbation, with mean $0$ and standard deviation $0.01$, to the value of $u$ and $v$ at each grid point. The equations are then numerically integrated until a steady state is reached.

\section{3D Simulations}
We use 3D models downloaded from https://free3d.com/. At each vertex, we calculate the principle curvatures and directions using the pymeshlab package. The inhibitors diffusion coefficient between neighbouring triangles of the surface, or cells, in fipy is modified as
\begin{equation}
    \frac{1}{2}\sum_v (d(\kappa^v_1) (\mathbf{t}^v_1 \cdot \mathbf{n})^2 + d(\kappa^v_2) (\mathbf{t}^v_2 \cdot \mathbf{n})^2)
\end{equation}
where we sum over the two vertices shared between the neighbouring cells, $k^v_i$ and $\mathbf{t}^v_i$ are the curvatures and principle directions of the vertices, $\mathbf{n}$ is the normal direction between the two cell centers, effectively giving a curvature to the interface which is average between the two vertices. Finally, we numerically integrate our equations until a steady state is reached.


\bibliography{refs}